\documentclass[sigconf]{acmart}
\usepackage{subcaption}

\setcopyright{rightsretained}

\acmConference[Preprint]{Arxiv Preprint (not submitted)}{January 2020}{}

\begin{document}

\title{Seeker or Avoider? User Modeling for Inspiration Deployment in Large-Scale Ideation}

\author{Maximilian Mackeprang}
\email{maximilian.mackeprang@fu-berlin.de}
\affiliation{%
  \institution{Human-Centered Computing, Freie
Universit\"at Berlin}
  \streetaddress{14195 Berlin, Germany}
}

\author{Kim Kern}
\email{kim.kern@fu-berlin.de}
\affiliation{%
  \institution{Human-Centered Computing, Freie
Universit\"at Berlin}
  \streetaddress{14195 Berlin, Germany}
}

\author{Thomas Hadler}
\email{thomas.hadler@fu-berlin.de}
\affiliation{%
  \institution{Human-Centered Computing, Freie
Universit\"at Berlin}
  \streetaddress{14195 Berlin, Germany}
}

\author{Claudia M\"uller-Birn}
\email{maximilian.mackeprang@fu-berlin.de}
\affiliation{%
  \institution{Human-Centered Computing, Freie
Universit\"at Berlin}
  \streetaddress{14195 Berlin, Germany}
}


%
\begin{abstract}
People react differently to inspirations shown to them during brainstorming. Existing research on large-scale ideation systems has investigated this phenomenon through aspects of timing, inspiration similarity and inspiration integration. However, these approaches do not address people's individual preferences. In the research presented, we aim to address this lack with regards to inspirations. In a first step, we conducted a co-located brainstorming study with 15 participants, which allowed us to differentiate two types of ideators: Inspiration seekers and inspiration avoiders. These insights informed the study design of the second step, where we propose a user model for classifying people depending on their ideator types, which was translated into a rule-based and a random forest-based classifier. We evaluated the validity of our user model by conducting an online experiment with 380 participants. The results confirmed our proposed ideator types, showing that, while seekers benefit from the availability of inspiration, avoiders were influenced negatively. The random forest classifier enabled us to differentiate people with a 73 \% accuracy after only three minutes of ideation. These insights show that the proposed ideator types are a promising user model for large-scale ideation. In future work, this distinction may help to design more personalized large-scale ideation systems that recommend inspirations adaptively.
\end{abstract}


\maketitle


\section{Introduction}
Generating ideas is a complex, multi-faceted process. People approach this process of idea generation, i.e. ideation, differently; they use distinct strategies or employ various methods \cite{garfield2001modifying}.
One established technique for improving creativity during ideation is providing people with others' ideas (subsequently called inspirations). The effect of others' ideas as an inspiration technique has led to the original proposal of \textit{brainstorming}~\cite{osborn1953applied}. The rise of digital media allowed the moving of co-located, small-scale brainstorming into large-scale online settings. In the latter, inspirations have become a critical aspect in the algorithmic support of brainstorming~\cite{siangliulue2015toward}. The ideation process itself is defined as follows: A problem or topic is provided as a challenge on an online ideation system. In a crowdsourced fashion, participants provide their ideas in individual time-bounded ideation sessions after reading the challenge. 
Related work has found that the effectiveness of inspirations depends on a person's cognitive state~\cite{siangliulue2015providing}, the semantic similarity of the inspirations provided~\cite{chan2017semantically} and the level of attention a person devotes to an inspiration~\cite{girotto2017effect}. However, when summarizing the existing approaches of providing inspirations in large-scale ideation systems, we can surmise that these approaches disregard individual preferences people might have when using inspirations during ideation. This gap in research led to our overarching goal: We envision that large-scale ideation systems consider individual user preferences (e.g. their cognitive state, the available type of inspirations and the particular context of ideation) in order to provide the most effective inspirations during ideation.

In the context of this work, we focus on one aspect in the context of user modeling and tackle the research question: Can we distinguish individual preferences towards inspirations in ideation session and how do these preferences influence the ideation outcome?
To approach this question, we conducted an exploratory pre-study, which allowed us to differentiate people who either seek or avoid inspirations during ideation. We translated this observation into an actionable user model and implemented the model in a classifier. 
We conducted an online experiment to validate our user model, consisting of two sessions. In the first session, participations were classified into one of the two ideator types (seeker or avoider) identified in the pre-study. Based on the type identified, we assign ideators to two conditions during the subsequent second, actual, ideation session. We then evaluated the impact of inspiration availability on the two types. Finally, we propose and evaluate a heuristic to dynamically detect inspiration seekers and avoiders in the context of a large-scale ideation session.

\section{Related Work}
Existing research in the research field of ideation can be divided into two main research streams. 
On the one hand, there is research from psychology, which analyzes mainly individual differences in group-brainstorming contexts. Garfield et al., for example, show that personality types influence the outcome of ideation~\cite{garfield2001modifying}. Even though these personality types might be a promising approach for modeling user preferences, the effect of inspiration has not yet been evaluated. Furthermore, Garfield's research is based on data obtained in a co-located group setting. Gamper et al. also conducted a study in the context of small scale brainstorming where they gave participants the chance to have their ideas critiqued and reviewed by other participants during a session. They show that in terms of this type of feedback, people can be classified into participants wanting feedback early, and others who deferred it to later~\cite{gamper2017sooner}. These insights can potentially be transferred to inspirations: Both feedback and inspiration are instances of external input during ideation. However, while inspiration is inspiring (divergent) input, feedback is normally viewed as restricting (convergent) input\cite{wang2017literature}.
On the other hand, there is an extensive body of work on inspirations in the context of large-scale ideation \cite{girotto2019crowdmuse,yu2016encouraging,chan2016ideagens,siangliulue2016ideahound,yu2016distributed,javadi2019experimental}. 
Siangliulue et al., for example, compared different timing mechanisms for providing inspirations in large-scale ideation systems~\cite{siangliulue2015providing}. The authors analyzed the impact of the inspiration mechanism in three formats: Inspirations are shown to the user (1) on demand, (2) on idle time, i.e. when the user is pausing, or (3) in fixed time intervals. However no user model was deployed in this research--individual preferences were thereby neglected.
In their research, Girotto et al.~\cite{girotto2019crowdmuse} proposed a user model. When users submit an idea, in their system, they have to choose one or two topical categories for it. This information is used to build a matrix model of categories to which each user has submitted ideas. This model can then be used to infer categories a user is likely to be fluent in (based on collaborative filtering techniques) and recommend inspirations according to this information. Although this approach introduced a user model, the model is based only on the categorization of ideas. There was no investigation of individual differences regarding how participants reacted to, or used the inspirations.
In summary, research in psychology highlights individual differences in ideation; however, this insight is evaluated only in co-located group settings. Research in the area of large-scale ideation focuses on either state modeling (timing) or categories during ideation; individual preferences have not been considered so far. This situation motivated our research on users' preferences when providing inspirations, which is presented in the following.

\section{Pre-Studies}

While related work investigated the impact of inspiration timing or individual preferences towards feedback, preferences towards inspirations have not yet been analyzed. We conducted an exploratory pre-study to better understand the role of individual preferences towards inspiration, i.e. how people react to and use inspirations. 

\subsection{Pre-Study in Co-located Ideation Setting}
The goals of the exploratory study were to identify potential individual preferences of participants and understand how the participants themselves assess their experiences.
As these qualitative insights are hard to obtain in a large-scale ideation context~\cite{heerwegh2009mode}, we decided to conduct three brainstorming sessions in a co-located group setting to get a more profound understanding about people's behavior during ideation. We attain a more detailed perspective on users' preferences by studying the situated practice of people during such co-located settings.
During the session, we collected a variety of data. Two of the authors directly observed the brainstorming sessions. At the end of each session, we asked participants to fill out a questionnaire that consisted of seven questions focusing on participants' individual experience. We asked participants, for example, when they used the provided inspirations, how they used the provided inspirations, and if the inspirations distracted or inspired them\footnote{The complete survey can be found at <anonymized>.}.
We used the \textit{brainwriting pool} method for the brainstorming sessions~\cite{schlicksupp1975grundlagen}. Participants in a brainwriting pool write down ideas on sheets of paper. These sheets are then placed in the middle of the table (the pool), where other participants can take and read them. We chose the brainwriting pool as it resembles many of the characteristics of large-scale ideation approaches. We recruited 15 participants via a mailing list, posters and personal networks. We conducted three sessions each with five people. The participants first received a short introduction to the method and brainstorming challenge (cp. \textit{Smart Coating} in Table~\ref{tab:ideation-challenges}) and then brainstormed ideas for 30 minutes. Afterwards, we provided the survey and the participants had 15 minutes to fill it out.

\begin{figure}[tb]
  \centering
  \includegraphics[width=0.6\linewidth]{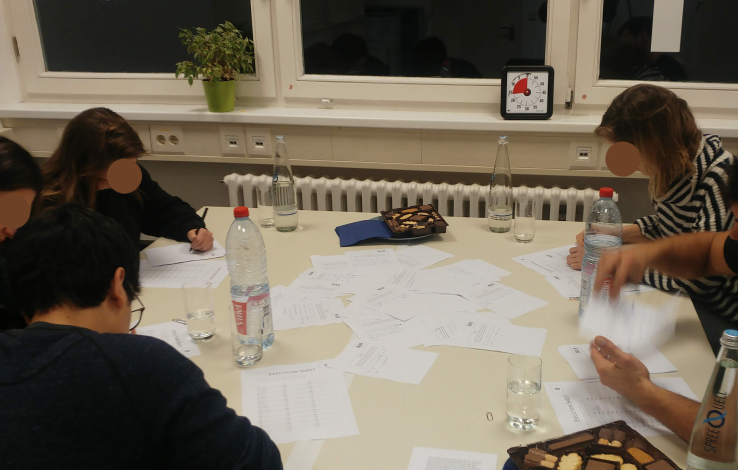}
  \caption{Study setup for the exploratory study.}
  \label{fig:exploratory-study-foto}
\end{figure}

\subsubsection{Results}

Overall, the 15 participants generated a total of 225 ideas and read 220 ideas for inspiration. Furthermore, all of the participants filled out the questionnaire. The questionnaires were analyzed using thematic analysis~\cite{peterson2017thematic} by one of the authors with a focus on how the participants perceived the study setup, its influence on their performance and differences in their behavior during the ideation session. This way, similar statements (e.g. P4: ``[\textit{others' ideas}] mostly inspired me because they showed very diverse use cases I was not thinking about'', P15:``Most of [\textit{others' ideas}] inspired me. I've noticed that my ideas were more diverse after reading others'.'') occurring multiple times were condensed into themes. By considering these themes, the completed questionnaires were re-read to ensure that all answers matching the themes are identified.
We derived three main themes from this qualitative analysis. These themes were \textit{social pressure} (P9: ``made me nervous how productive they seemed to be''), \textit{inspiration integration} (e.g. annotating how participants' ideas re-used inspiration aspects) and \textit{attitude towards inspirations} (P13:``some of the ideas helped me think in a different way'').
For the context of this work, we subsequently focus on the latter: Ideators' attitude about inspirations. We refer to <anonymized> for other results and a more detailed description of the methods used.
In the overall attitude towards others' ideas as source of inspiration, we observed two main behaviors: One group of participants stated the beneficial aspect of others' ideas (P2: ``[\textit{other participant's inspirations}] showed me which areas were missing'', P10: ``I looked at them to change the area of application I was into''). Another group of participants reported that they were distracted by the inspirations (P7: ``When the ideas were way out of my imagination it distracted me a bit.''). One participant even deliberately disregarded using inspiration (P1: ``No, I was too set in my ways. Didn't really want to change my methodology.''). 
Based on the results of this study, we summarized our understanding of preferences towards inspiration in two archetypal users (subsequently called ideator types). We hypothesize that the two types exemplify different strategies to come up with ideas:

\begin{description}
    \item \textbf{Inspiration Seekers} are actively looking for inspirations during an ideation process and derive their ideation strategies from them.
    \item \textbf{Inspiration Avoiders} feel distracted by inspirations. They follow their own ideation strategies to come up with more or better ideas.
\end{description}

However, these two ideator types are derived from qualitative insights and obtained in the context of co-located group settings. In order to further substantiate the definition, we analyzed data from large-scale ideation sessions collected in previous research. In the next section, we describe the results of this data analysis.

\subsection{Deriving a Heuristic for Determining Seekers and Avoiders}


We conducted an exploratory data analysis to transfer our preliminary model into the context of large-scale ideation. The goal of this data analysis was to compare the insights about the ideator types found in the co-located setting with existing data obtained in a large-scale ideation context. We wanted to define the different ideator types in terms of data available in large-scale ideation (e.g. tracking data).
We used tracking data of three studies\footnote{The data is published anonymously at \url{https://osf.io/7wjya/?view_only=0ba9e138d22e414abd8b868ed594e93e}}. The data was obtained by using a web-based prototype similar to the one outlined in Section~\ref{sec:interface-description}. In these ideation sessions, participants were presented with a challenge and got a fixed amount of time to write ideas and submit them via the user interface. Furthermore, they were shown a button that allowed them to request inspirations\footnote{The inspirations are based on others' ideas.}. Overall, the data comprised 193 individual ideation sessions from three challenges (cp. Table~\ref{tab:data-analysis-input-datasets}). Each participant completed one session for one of the data sets.
\begin{table}[htb]
\caption{Data sets (DS) used for validating the seeker / avoider model. Data sets varied in their number of participants (each participant completed one session for one data set), length of the sessions and input challenge (see Table~\ref{tab:ideation-challenges} for a full description of the challenges).}
  \label{tab:data-analysis-input-datasets}
  \begin{tabular}{llll}
    \toprule
    Name & Sessions & Length (min) & Challenge \\
    \midrule
    DS1 & 89 & 15 & Smart Coating \\
    DS2 & 30 & 25 & Smart Coating \\
    DS3 & 74 & 20 & Bionic Radar \\
  \bottomrule
\end{tabular}
\end{table}
The data sets (DS1, DS2, DS3) consisted of timestamped events for all sessions, namely session start, session end, whether the tab with the application was focused, idea submissions and inspiration requests. We chose inspiration requests as the most relevant metric for a potential operationalization of the user model described previously. We decided to segment the participants into how often they requested inspiration to find the two approaches within the session data.
Furthermore, we disregarded all participants with fewer than three idea submitted (similar to studies in related work~\cite{siangliulue2015providing}) which resulted into 20 filtered sessions (10 for DS1, 1 for DS2 and 9 for DS3).

%

\subsubsection{Results}

\begin{figure}[htb]
  \centering
  \includegraphics[width=1\linewidth]{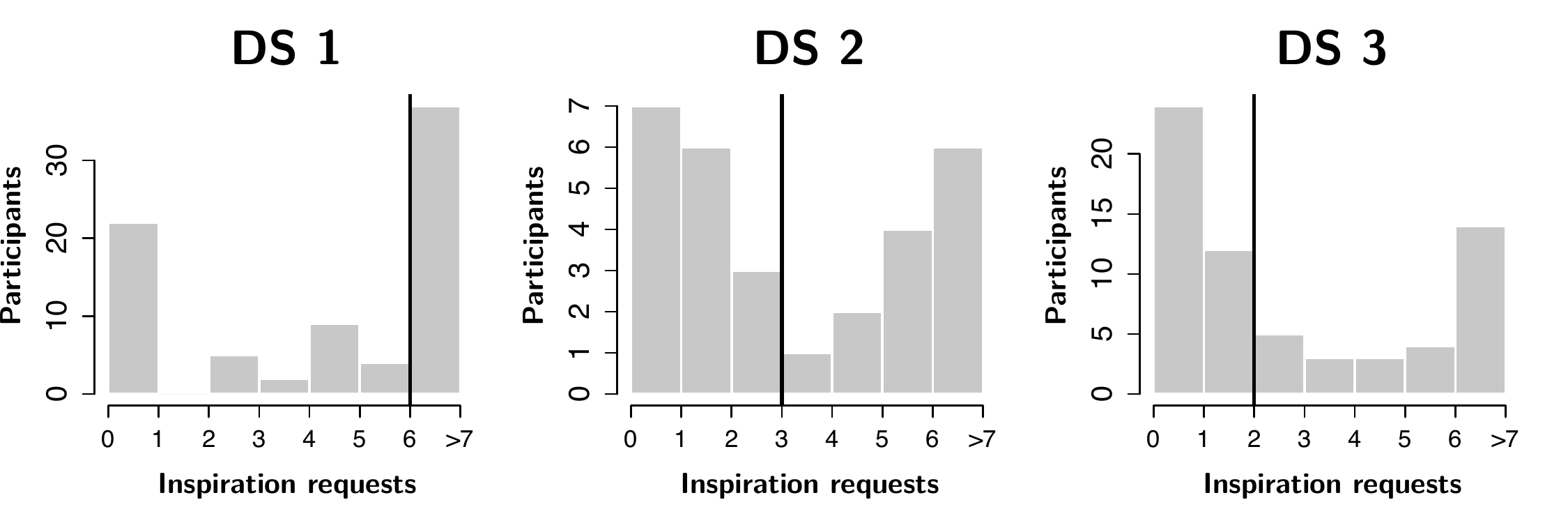}
  \caption{Distribution of inspiration requests over all participants for all data sets. The vertical line represents the median of inspiration requests. Participants are shown in absolute numbers.}
  \label{fig:inspiration-request-distribution}
\end{figure}

We show the distribution of inspiration requests in each data set in Figure~\ref{fig:inspiration-request-distribution}. We summarized users that requested inspirations for more than 8 times in the last bin to remove outliers. The participants requested inspiration up to 40 times in DS1, 20 times in DS2 and 15 times in DS3. The median number of inspiration requests were six for DS1, three for DS2, two for DS3.

One goal for the data analysis was to develop a model of seekers and avoiders based on the patterns in the data sets.
When visualizing the inspiration request distributions, we found that inspiration requests formed an almost bimodal distribution. However, there were some outliers in inspiration requests, where people requested inspiration heavily towards the end of the session. We hypothesize that these requests happened out of idleness and not with the intention of getting inspiration. As a first classification rule we, therefore, chose the median number of requests over all participants as the cut-off point between seekers and avoiders. We opted for the median because it is not as susceptible to outliers as the mean. When using the median as the cut-off point, we found that participants having five inspiration requests would still be classified as avoiders for DS1. Based on the insights from the co-located study, we decided to change our definition so that only people with one inspiration request at most were classified as avoiders. It turned out that the request was often conducted right at the beginning of the session in sessions with only one inspiration request. We assume that the participants in these sessions represent inspiration avoiders, because participants supposedly tested the inspiration button once and discarded it completely later during ideation. We, therefore, chose to expand our avoider definition to include participants with one request.
Based on these insights, we defined seekers and avoiders more precisely:
\label{sec:ideator-type-definitions}
\begin{description}
    \item If participants request more than the median number of inspiration requests, we assign them to the group of \textit{Inspiration Seeker}.
    \item If participants request inspiration at most once during the ideation session, we assign them to the group of \textit{Inspiration Avoider}.
\end{description}

Based on these definitions, we classified 80 participants as seekers and 51 as avoiders. We assigned all participants that fall in between these categories to the group \textit{Undetermined}. 
The rule-based definition of seekers and avoiders allowed us to classify existing sessions based on the complete information about the sessions. However, to understand how the ideator type impacts users we need typed participants to test with. We, therefore, analyzed partial data from the sessions to develop an understanding on when to classify participants.

\subsection{Developing a Dynamic Classification}
\label{sec:dynamic-classifciation}
The results of the visualization of the distribution of inspiration requests (see Figure \ref{fig:inspiration-request-distribution}) confirmed our hypothesis of the existence of ideator types and provided us with a rule-based definition based on finished sessions and ideation challenges.
We chose to conduct an online experiment to test the effects of ideator types in a large-scale ideation context. We decided to classify the participants of the experiment by conducting a classification session. However, one open question when choosing this approach is: How long should a classification session be? We conducted an analysis of partial data from the data sets to answer this question, which we present in the following.
The analysis was based on the ideator types we assigned to participants using the rule-based classification.
We divided the sessions into inspiration sequences to develop a feeling for how long a classification session should be.
An inspiration sequence describes the number of inspiration requests in bins of 60 seconds each.
We applied the rule-based classification on partial sequences and then compared the results with the label at the end of the session to find out a good length for a classification session.


\subsubsection{Results}
\begin{figure}[tb]
  \centering
  \includegraphics[width=1\linewidth]{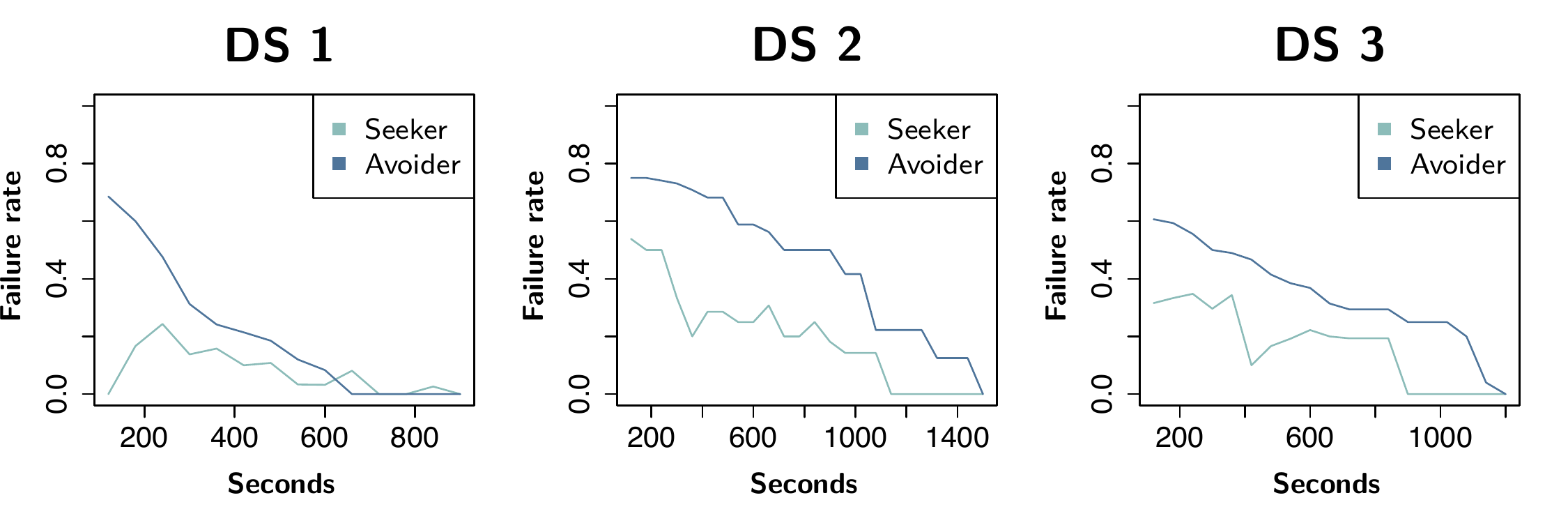}
  \caption{Ratio of incorrectly classified seekers/avoiders (failure rate). For each point in time, the classified ideator types are
compared to the results after the whole study duration.}
  \label{fig:failure-rates}
\end{figure}

We show the failure rates of applying this model in Figure~\ref{fig:failure-rates}. The number of incorrectly classified ideators decreases over time compared to the classifications at the very end of each session (e.g., in DS1 60\% of the avoiders were classified correctly after 300 s). 



\subsection{Discussion}
In the qualitative analysis of the co-located brainstorming setting, we found that some participants relied heavily on other participants' inspiring ideas, while others were either confused by inspirations or actively rejected them. The qualitative feedback from the study participants allowed us to get more insights into the characteristics of the participants. It turned out that avoiders are more confident about the task of ideation and their idea generation strategies, which we assume, increases their fluency, i.e. number of ideas generated during a session.  We conducted a data analysis of existing ideation data to validate these insights in the context of a large-scale online settings. 
We found in our analysis that the classification errors for avoiders decreased over time. Seekers, on average, could be classified faster (e.g. in DS1, 70\% of seekers are classified correctly after 300 s). The failure rate analysis in Figure \ref{fig:failure-rates} shows that at 10 min, the classification for both seekers and avoiders is better than by chance for all three data sets analyzed. Furthermore, the failure rate seems to decrease non linearly for DS1 and DS2. We decided to investigate this insight further, since an early classification of seekers/avoiders can enable an adaptive recommendation of inspirations in large-scale online ideation.
The results of the data analysis allowed the definition of a user model for inspiration seekers and avoiders. We operationalized this distinction based on the number of inspiration requests they submit during an ideation session.
We defined the following research questions to evaluate this user model:
\label{sec:research-questions}
\begin{description}
  \item $RQ_1$.  To what extend does ideator type (based on the rule-based classification) depend on the challenge?
 \item $RQ_2$. How do ideation metrics, such as fluency, differ between ideator types?
  \item $RQ_3$. To what extent are ideator types impacted by the availability of inspiration?
\item $RQ_4$. Are we able to predict the ideator type without full session data?
 \end{description}

We carried out an online experiment using Amazon Mechanical Turk (MTurk) to investigate these questions. The study setup and results are described next.

\begin{table}[htb]
\caption{Ideation challenges used in the studies.}
  \label{tab:ideation-challenges}
  \begin{tabular}{ll p{3cm}}
    \toprule
    Name & Used In & Description \\
    \midrule
    Smart Coating & Pre-Study & \small{Imagine you could have a coating that could turn every surface into a touch display. Brainstorm cool products, systems, gadgets or services that could be built with it.} \\
    Bionic Radar & Session A & \small{A technology can perceive the movement of an object, such as humans, living beings or objects,
like a bat. Once remembered, the technology can subsequently recognize it (...) The technology is approximately hand-sized and can be used anywhere.} \\
    Fabric Display & Session B & \small{Imagine there was a touch-sensitive ``fabric display'' that could render high resolution images and videos on any fabric through a penny-sized connector \cite{siangliulue2015providing}}. \\
  \bottomrule
\end{tabular}
\end{table}

\section{Study}
The qualitative feedback from the co-located ideation sessions and the preliminary data analysis from previous studies led us to the hypothesis that a classification of ideators exists relying on inspiration (seekers) and ideators distracted or annoyed by it (avoiders). However, this hypothesis is based only on qualitative findings. 
Furthermore, the existence of these ideator types says nothing about their effect on brainstorming or inspiration deployment. We conducted an online experiment to systematically evaluate the differences of ideator types in large-scale ideation. 
The goals of this experiment were, firstly, to validate the user model proposed, secondly, to test whether the model has relevant impact on user interface configuration and, thirdly, to see if we can classify ideators heuristically, even without having data on a full ideation session.


\subsection{Methodology}
We first needed to distinguish participants within our experiment to understand the impact of the model systematically. We, therefore, conducted a first ideation session as a classification-session (session A). Based on the inspiration requests in this session, we classified participants into the types described above. We then asked them to do another ideation session (session B). We collected data about ideation metrics in the second session, based on the classification. We, furthermore, split participants into two conditions (see below) to understand the impact of the ideator types as a user model in large-scale ideation. We decided to conduct both sessions in one Human Intelligence Task (HIT), because we did not have to recruit the people again but could allow participants to immediately continue working (to avoid low retention rates).

\begin{figure}[htb]
  \centering
  \includegraphics[width=\linewidth]{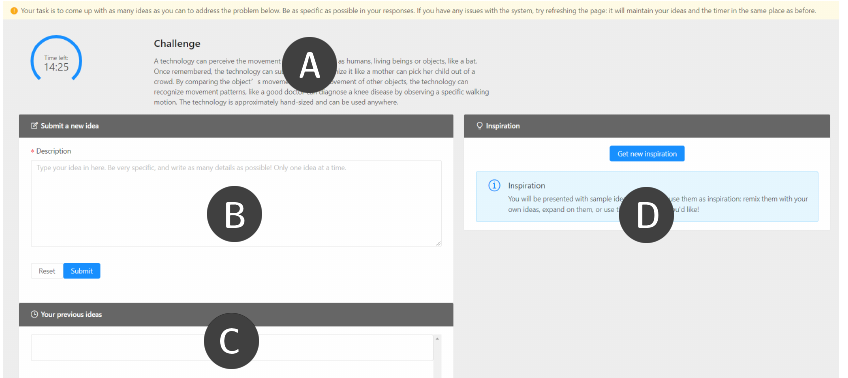}
  \caption{The brainstorming user interface, showing the challenge panel (A) showing the current ideation challenge, the idea input panel (B) where users type their ideas, an idea history list (C) showing all of the users' previous ideas and an inspiration button (D). In the \textit{baseline} condition, the inspiration button was not shown.}
\label{fig:interface}
\end{figure}

\subsubsection{Study Design}
\label{sec:interface-description}
The study consisted of two sessions. The first session used the same software for all participants. The software (as shown in Figure \ref{fig:interface}) consisted of a challenge description panel (A), an idea input panel (B), where participants could enter ideas, an idea history (C) and an inspiration button (D).

The data obtained in the classification session was used to assign a type to the participant. If a participant issued zero or one inspiration request, we classified them as an avoider. As we did not have the median number of inspirations available, we chose a cut-off of four to identify a participant as a seeker. This cut-off was chosen qualitatively after the exploratory data analysis. It differed from the median calculated after the experiment (5) by one.
This means that participants that requested between two and four inspirations were classified as \textit{undetermined}. We classified participants submitting three or fewer ideas as \textit{unmotivated}.
We used a 2x2 factorial design for the second session to combine ideator types with two different conditions. Participants in the \textit{on-demand} conditions were shown an \texttt{[inspire me]} button. The button was not shown for participants in the \textit{baseline} condition.

\subsubsection{Challenge}
Participants started the study with a short text introduction describing the brainstorming challenge and the user interface. Participants were then shown the main brainstorming app (as shown in Figure \ref{fig:interface}) and were asked to brainstorm ideas for 10 min (session A). The timing of session A was chosen based on the results from Section \ref{sec:dynamic-classifciation}. The challenge for this session was \textit{bionic radar}, with the prompt listed in Table \ref{tab:ideation-challenges}. 
People with a classification of seeker or avoider moved on to the second session (session B). The other participants were directed to the survey. In session B, participants had to brainstorm for 15 min on the challenge \textit{fabric display} (see Table \ref{tab:ideation-challenges}). We chose to change the challenge between session A and B to evaluate whether seeker / avoider is a feature of the participant or the challenge at hand. All participants filled out a user experience survey at the end.

\subsubsection{Inspiration Mechanism}
The inspiration mechanism implemented was an \texttt{[inspire me]} button that participants could click on when they were stuck. This \textit{pull}-approach to inspiration is based on related work comparing different timings of inspiration deployment \cite{siangliulue2015providing,chan2017semantically}.
When the inspiration button was clicked on, the participant received one inspiring idea. In order to obtain the inspirations, 200 ideas for both challenges from previous studies were manually rated for quality (in terms of novelty and value) by two of the authors. The inspirational ideas were sorted by the sum of novelty and value and then sorted (each inspiration button click returned the next idea, in descending order). This approach ensured that only high-quality ideas were shown as inspirations (as suggested by related work \cite{siangliulue2015providing}). Furthermore, this approach ensured that inspirations were equal for all participants.

\subsubsection{Participants}
We recruited 380 participants using Amazon Mechanical Turk. We limited participation to U.S. workers with at least 1,000 HITS and an approval rate $>95\%$. Participants received \$ 4 if they completed one session and \$ 7 if they completed both sessions (\textasciitilde \$12/h).

\subsection{Study Part 1}
Our focus in the first part of the study, was on the ideation process: By conducting the classification session, our goal was to analyze the behavior of seekers and avoiders separately, and understand how the proposed model translates into ideation metrics. 
We were able to analyze whether the availability of inspiration impacts avoiders differently than seekers by having the availability of inspiration as a condition in the second study. Lastly, we wanted to find out if the avoider/seeker classification is dependent on the challenge or the participant by having two sessions with different challenges. We defined the following hypotheses to test these goals:

\begin{description}
    \item $H_1$: The seeker / avoider distinction is a characteristic of the user and not the challenge. Therefore, participants classified as avoiders in the first session will not request inspiration in the second session.
    \item $H_2$: The availability of inspiration distracts avoiders, leading to a lower quality of their ideas in the \textit{on-demand} condition.
\end{description}


\subsubsection{Measures and Analysis}
We chose to employ standard ideation metrics, as used in related work \cite{runco2012standard}, to analyse the effects of ideator types and the availability of inspiration. Firstly, we measured \textit{Fluency}: the number of ideas generated in a session. Furthermore, we tracked the number of times, a participant clicked on the inspiration button (in session A and in the \textit{on-demand} condition of session B).
We employed \textit{novelty} and \textit{value} ratings from external crowdworkers to analyse the effect on ideation outcome. We collected ratings for all ideas generated in session B by asking workers to rate novelty and value on a five-point Likert scale for a batch of fifteen ideas per HIT. Ratings were obtained redundantly, with at least three ratings per idea (some of the ideas were rated more often because assigned batches were returned unfinished by workers, leading to a non-uniform distribution of batches). We normalized the ratings for each batch to account for systemic bias in the ratings (e.g. a worker rating all ideas in their batch at 5 (\textit{very good})). We decided to test the maximum novelty and value per participant. This decision was motivated by the overarching goal of brainstorming: The statistical chance of high quality ideas is increased by contributing large quantities of ideas. Therefore, we compared only the ideas with the highest novelty and highest value within their session to measure the effect of inspiration availability on ideators.

\subsubsection{Results}
A total of 380 MTurk crowd workers participated in the experiment. Of these, 134 participants continued with session B, from whom 81 sessions were considered usable. We filtered 41 sessions that were placed in a faulty condition, due to a software error. The remaining 12 participants were sorted out because they had either submitted only non-sense ideas (e.g. ``this a machine to know underwater scenery that makes the more deeps goes that due to make the changes'') or completely misunderstood the challenge. The other 246 participants only completed session A and the survey, because they were either not classified as a seeker or avoider, or the target number of participants per condition had already been reached (e.g. a participant who was classified as a seeker did not complete session B if we already had obtained 40 observations classified as seekers).
Out of the 380 participants, based on the inspiration requests, we classified 201 as seekers, 59 as avoiders, 83 as undetermined and 37 as unmotivated. In session A, the participants requested inspiration 3,018 times overall, with a median number of inspiration requests of five.
\begin{figure}[htb]
  \centering
  \includegraphics[width=0.8\linewidth]{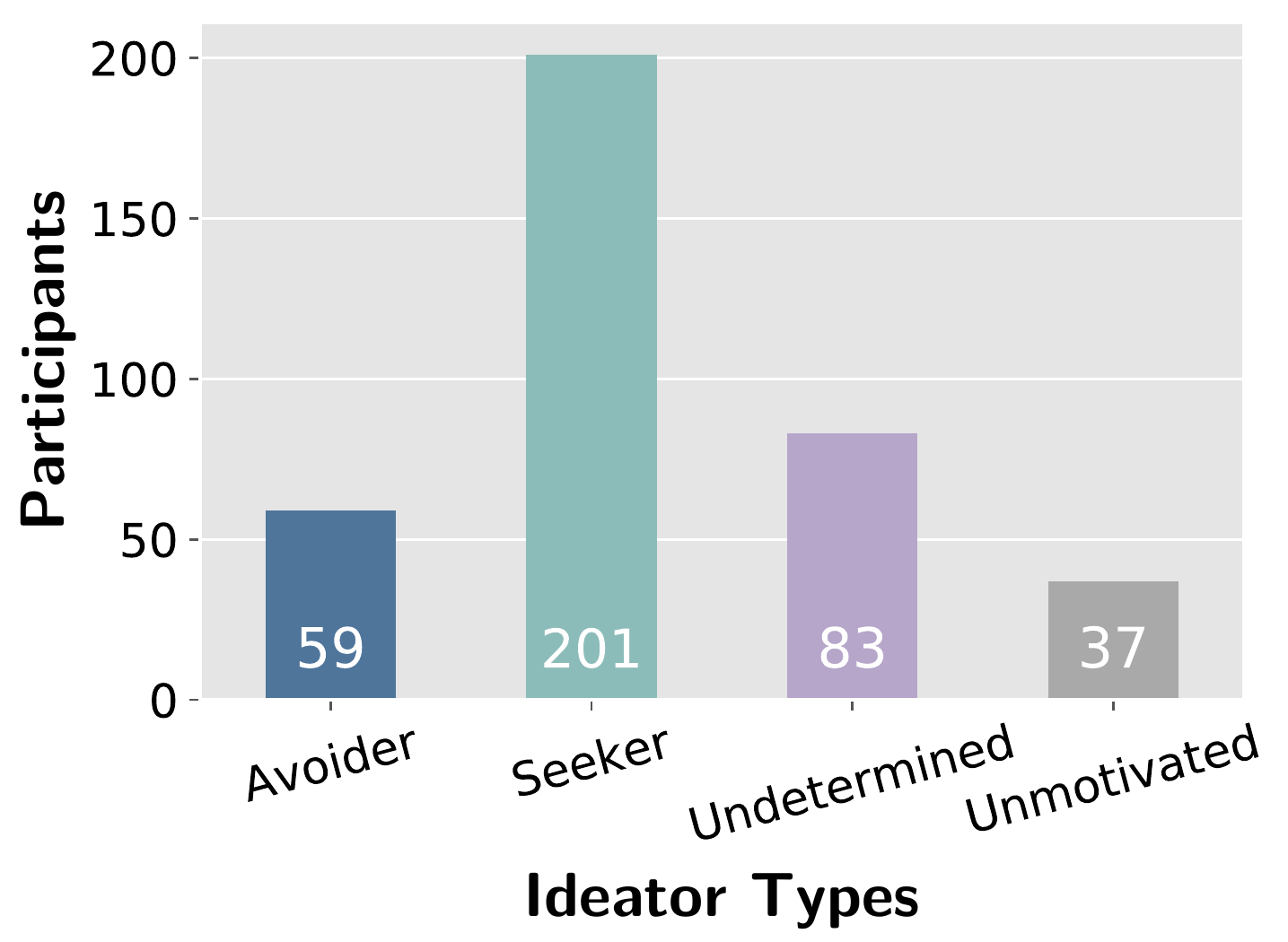}
  \caption{Number of participants classified into types after session A.}
  \label{fig:type-numbers}
\end{figure}



\begin{figure}[htb]
  \centering
  \includegraphics[width=0.6\linewidth]{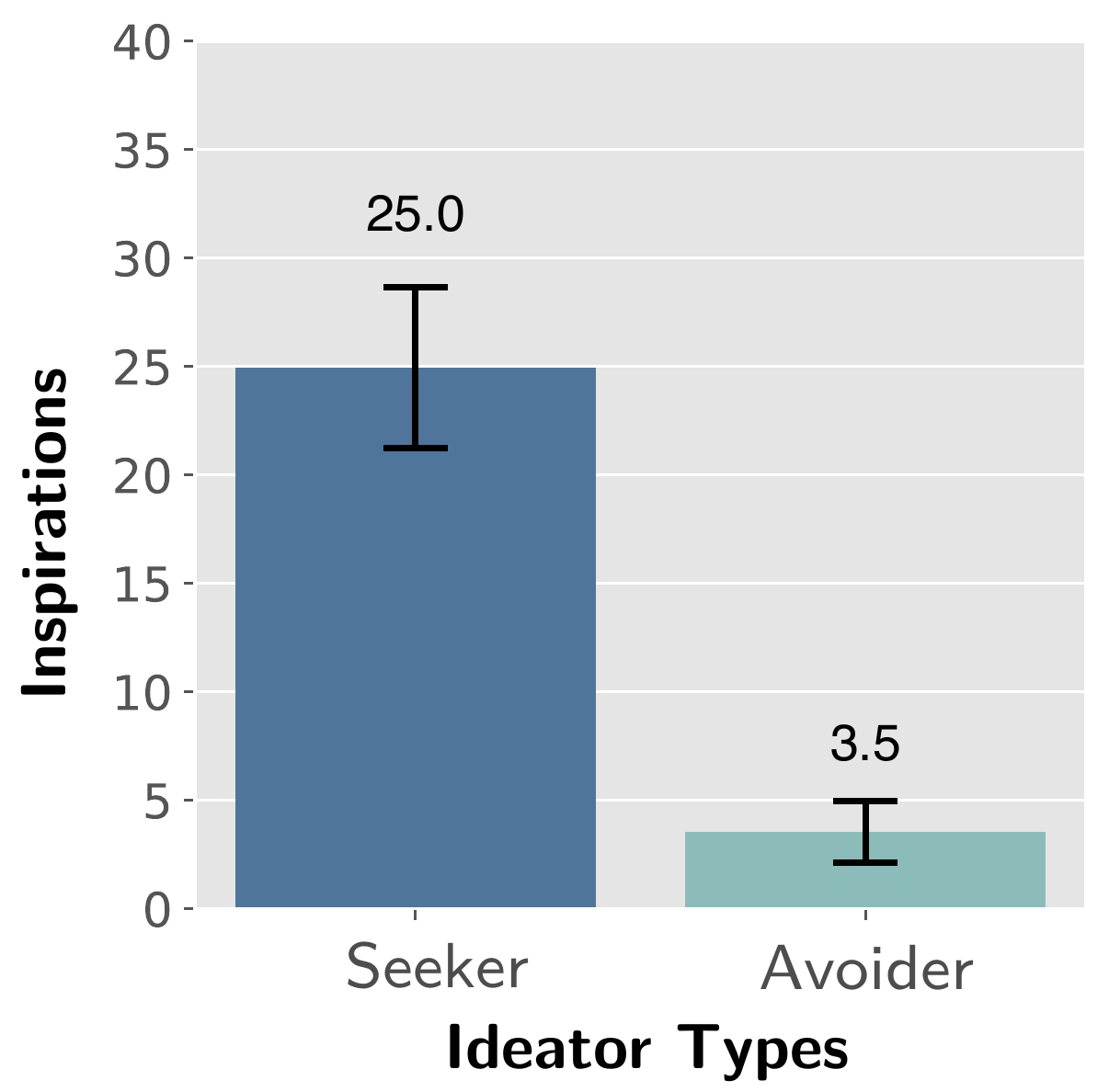}
  \caption{Number of inspirations requested by participants in the \textit{on-demand} condition of Session B, segmented by label. The error bars represent the standard error.}
  \label{fig:inspiration-requests-session-b}
\end{figure}

\paragraph{Ideator Types}
To understand whether the ideator type depends on the challenge, we compared the behavior of the participants in session A to the participants in the \textit{on-demand} condition of session B. Intuitively, participants that were classified as avoiders in the first session should not use the inspiration button in session B.
Figure \ref{fig:inspiration-requests-session-b} shows the inspiration requests of seekers and avoiders in session B. To test $H_1$ we deployed a linear model predicting the number of inspiration requests for ideators in the on-demand condition. The model shows a highly significant difference between seekers and avoiders $(p < 0.001, r = 0.412, r^2= 0.402, \text{estimate} = 21.402)$.
We furthermore used the rule-based classification based on the tracking data of session B in the on-demand condition. Table \ref{tab:classification-movement} shows the participants' labels after session A and then, if the label changed, how many participants were re-classified in session B. We found out that all the participants classified as seekers in session A stayed in that classification. On the avoider side, 11 of the 20 avoiders would be classified as avoiders again. The rest was classified as undetermined (5), seekers (3) or unmotivated (1).

\begin{table}[htb]
\caption{Classifications after Session A and Session B}
  \label{tab:classification-movement}
    \begin{tabular}{llr}
    Session A & Session B    & \# Participants \\
    \midrule
    Seeker    & Seeker       & 21            \\
    Avoider   & Avoider      & 11            \\
    Avoider   & Seeker       & 3             \\
    Avoider   & Undetermined & 5             \\
    Avoider   & Unmotivated  & 1            
    \end{tabular}
\end{table}

\paragraph{Impact of Ideator Type and Condition on Ideation Metrics}
We used linear regression models to test $H_2$.
All linear models were tested against their assumptions according to Field et al. \cite{field2012discovering}. The data were log-transformed for idea submits to ensure a normal distribution of model residuals as suggested by the Box-Cox test \cite{box1964analysis}. A probable cause for the positively skewed distribution is that ideators with a low number of submits were sorted out as unmotivated. The contrasts were coded as described by Schad et al. \cite{schad2018capitalize}: The ideator type was coded as a sliding difference contrast, and the condition was coded as a custom contrast with on-demand vs. baseline so that the estimates can be interpreted.

\begin{figure}[htb]
\centering
\begin{subfigure}[c]{1\linewidth}
   \includegraphics[width=\linewidth]{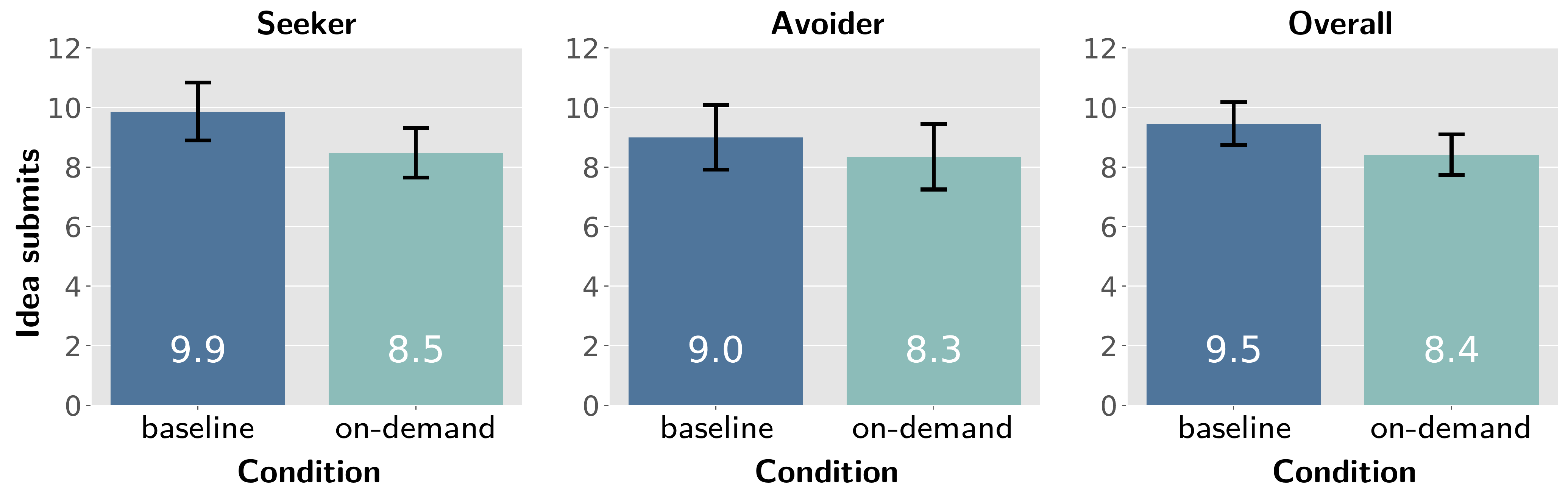}
   \subcaption{Fluency}
   \label{fig:fluency} 
\end{subfigure}
\caption{Fluency for session B per ideator type and condition. Overall shows seekers and avoiders together. The error bars represent the standard error.}
\end{figure}

\begin{figure}[htb]
\centering
\begin{subfigure}[c]{1\linewidth}
   \includegraphics[width=\linewidth]{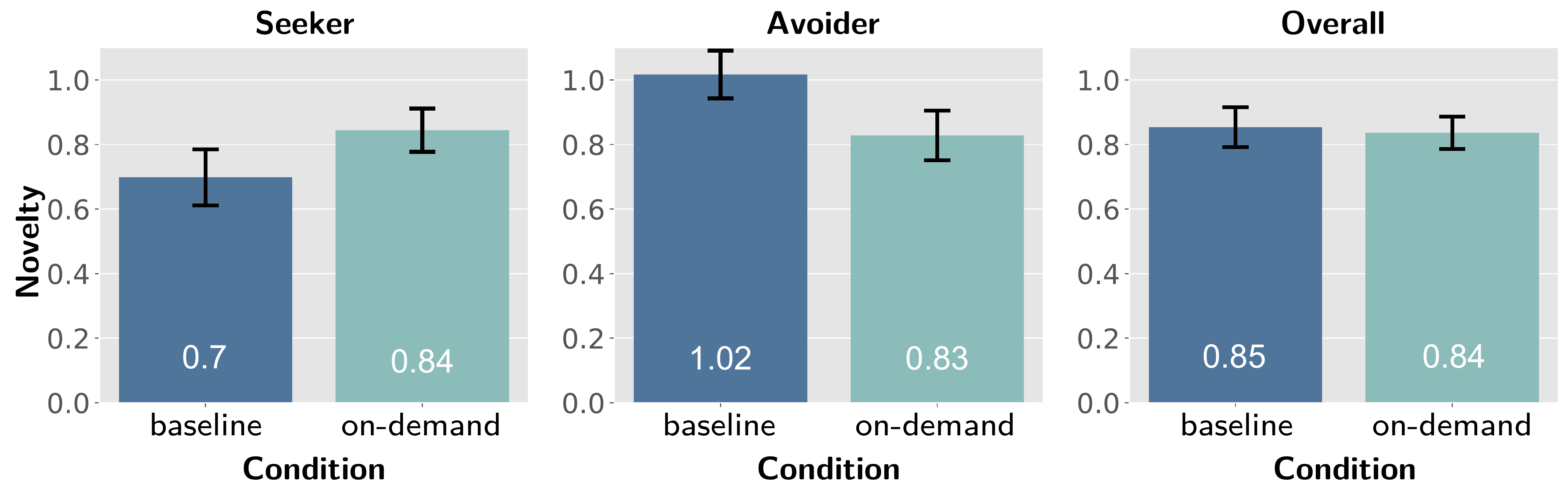}
   \subcaption{Max Novelty}
   \label{fig:max-novelty} 
\end{subfigure}
\begin{subfigure}[c]{1\linewidth}
   \includegraphics[width=\linewidth]{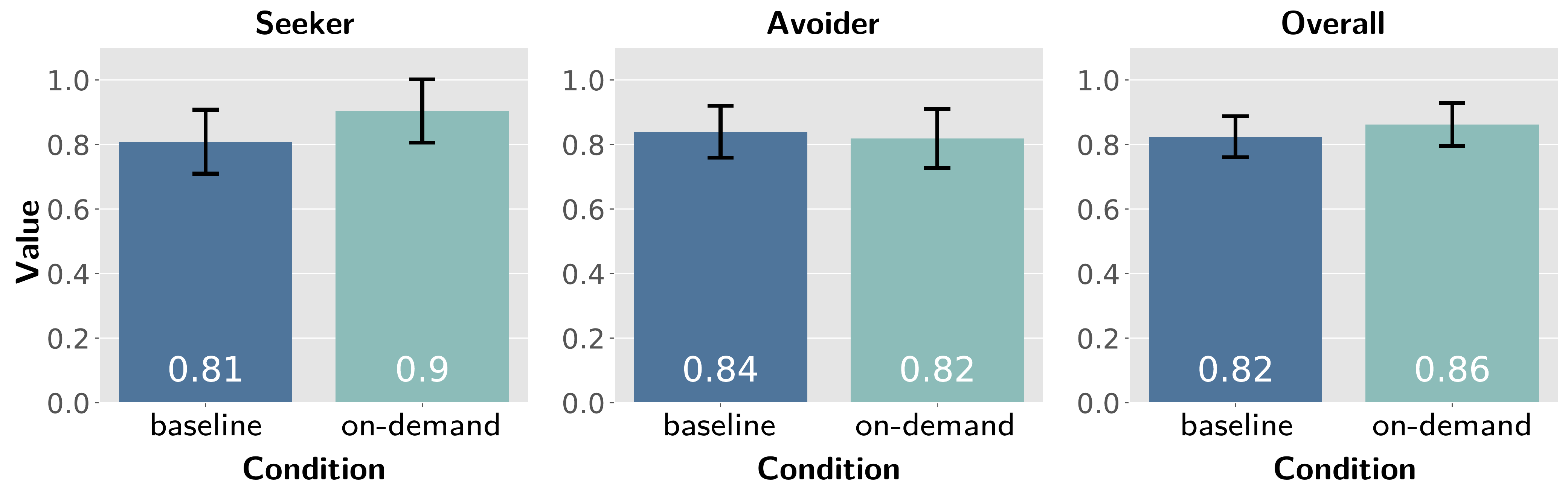}
   \subcaption{Max Value}
   \label{fig:max-value} 
\end{subfigure}
\caption{Maximum novelty/value ratings per ideator type and condition. Maximum novelty/value is defined as the idea with the highest novelty/value rating within a session. The error bars represent the standard error.}
\end{figure}

\begin{figure}[htb]
\begin{subfigure}[c]{0.49\linewidth}
\includegraphics[width=\linewidth]{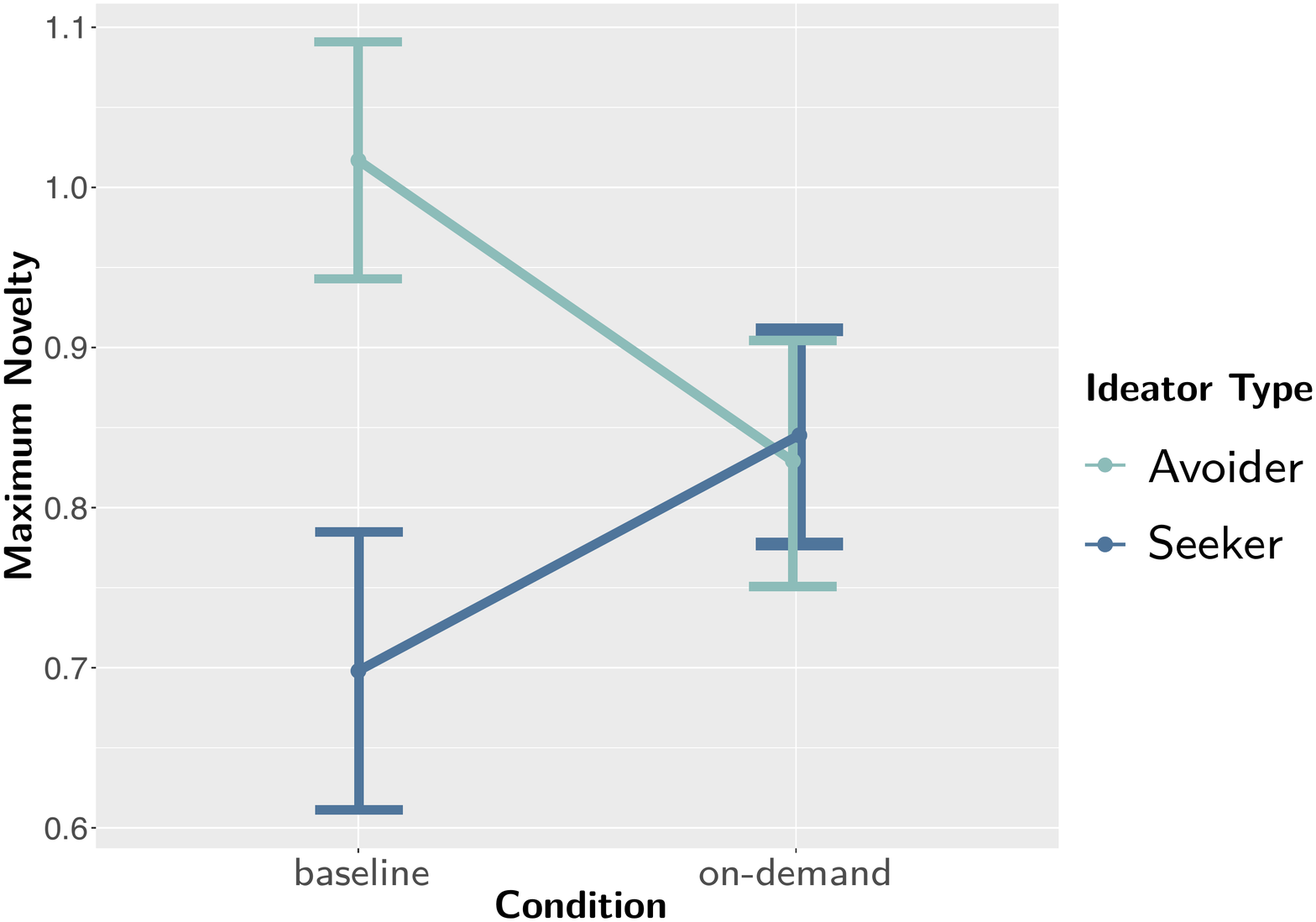}
\subcaption{Max Novelty}
\label{fig:interaction-effect-max-novelty}
\end{subfigure}
\begin{subfigure}[c]{0.49\linewidth}
\includegraphics[width=\linewidth]{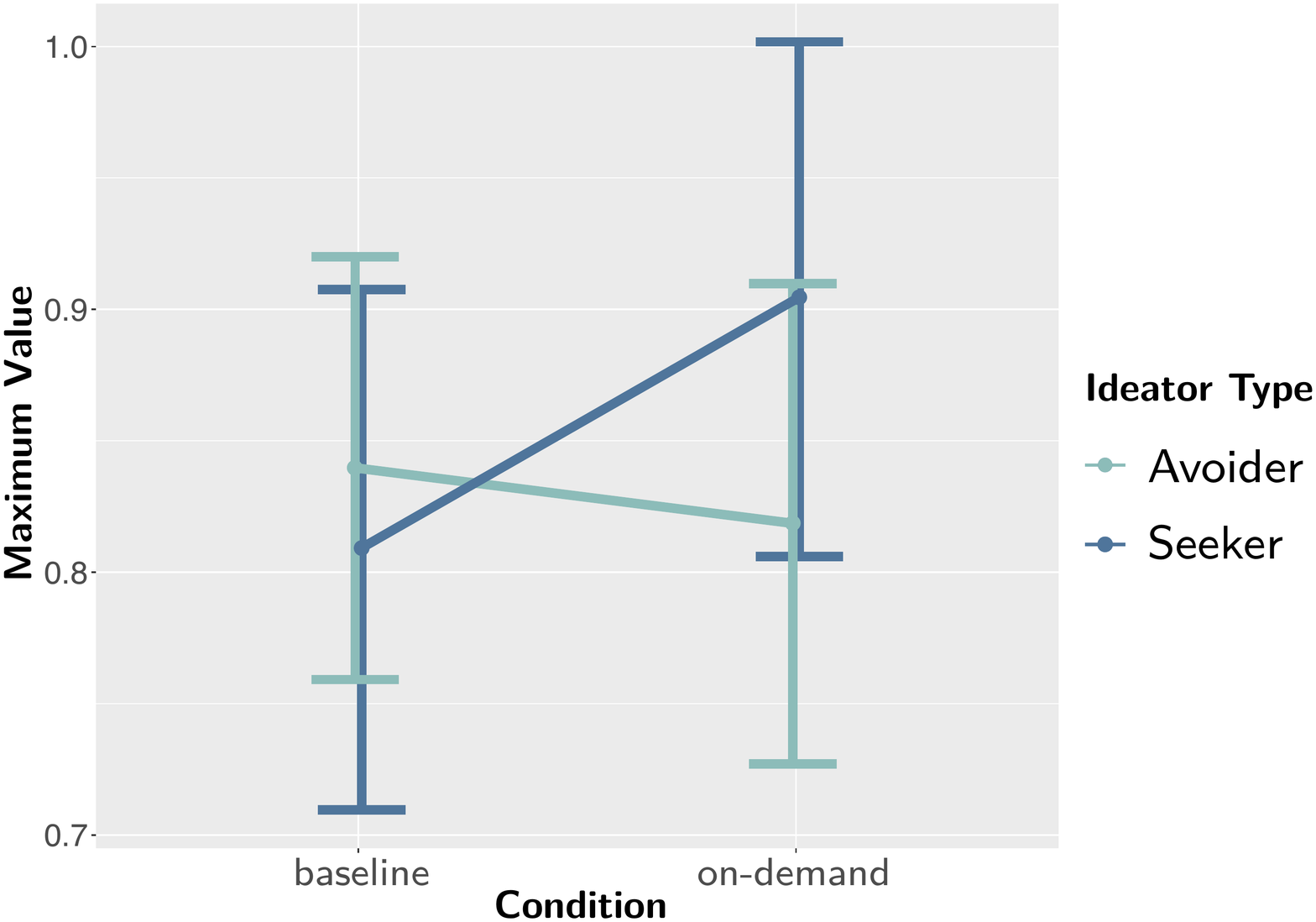}
\subcaption{Max Value}
\label{fig:interaction-effect-max-value}
\end{subfigure}
\caption{Interaction Effects for conditions and ideator types}
\end{figure}

We collected ideation metrics (fluency, novelty, value) for session B and segmented them by condition. Figure \ref{fig:fluency} shows the fluency results for types and conditions in session B.
There were no main or interaction effects in the multiple linear regression on fluency between ideator types and conditions $(p=0.964, r^2= 0.040, \text{estimate} = -0.15)$. Fluency was slightly lower for avoiders than seekers, albeit non-significant.
In the multiple linear regression on maximum novelty between ideator types and conditions, there was a significant main effect for ideator types ($p < 0.001$) and significant interaction effects between ideator types and conditions ($p = 0.043$) see Table \ref{tab:max-novelty-model}. On average, an avoider's most novel idea had a higher rating than a
seeker's most novel idea (difference = 0.11), see figure \ref{fig:max-novelty}. But more importantly,
the interaction effect significantly shows, that avoiders increase their novelty ratings when moving from \textit{on-demand} to \textit{baseline}, whereas it is the other way around for seekers, see Figure \ref{fig:interaction-effect-max-novelty}.

\begin{table}[htb]
\caption{Multiple linear regression model predicting maximum novelty based on ideator type and condition.}
\label{tab:max-novelty-model}
\begin{tabular}{lrrrr}
\toprule
\multicolumn{1}{c}{Predictors}                                                            & \multicolumn{1}{c}{Estimates} & \multicolumn{1}{c}{SE} & \multicolumn{1}{c}{Statistic} & \multicolumn{1}{c}{p} \\ \midrule
Grand Mean                                                                                & 0.83                          & 0.03                   & 24.61                         & \textless{}0.001      \\
Seeker vs. Avoider                                                                        & -0.11                         & 0.07                   & -1.68                         & 0.096                 \\
on-demand vs. baseline                                                                    & -0.02                         & 0.08                   & -0.26                         & 0.793                 \\
\begin{tabular}[c]{@{}l@{}}Seeker vs. Avoider :\\ on-demand vs. baseline\end{tabular} & 0.34                          & 0.16                   & 2.04                          & 0.043                 \\
Observations                                                                              & 122                           &                        &                               &                       \\
$R^2$ / $adjusted R^2$                                                                    & 0.067 / 0.026                 &                        &                               &                       \\ \bottomrule
\end{tabular}
\end{table}

There were no main or interaction effects in the multiple linear regression on maximum value between ideator types and conditions, see Figure \ref{fig:interaction-effect-max-value}. Differences in maximum value were marginal, see Figure \ref{fig:max-value}.

\subsection{Study Part 2}
The second goal of the study was to find out whether we can predict ideator types based on incomplete sessions. This objective was motivated by the results of the failure rate analysis in the pre-studies (cf Section \ref{sec:dynamic-classifciation}) and the vision of an personalized adaptive inspiration recommender system.
The input data are the number of inspiration requests of the sessions summed in bins over equal amounts of time. We used both sessions as the input data. Session A contains 343 sessions (we filtered 37 unmotivated participants from the original 380 samples) and took 10 min. Session B in the \textit{on-demand} condition contains 41 samples and took 15 min.
Since we want to determine a user's ideator type before the session ends, we restrict the input data's end point in time, looking at only the first x minutes of the session. The output is a classification of the participant as seeker or avoider (based on the entire session\textquotesingle s data). Table \ref{tab:example-dataset} shows a sample session, where the user requested 0 inspirations in the first bin, then 1 inspiration in the second bin. This session was labelled as an avoider by the procedure detailed in Section \ref{sec:ideator-type-definitions}.
\begin{table}[htb]
\caption{Example 10-minute data set with a 15 seconds bin-size which gives us 40 bins.}
\label{tab:example-dataset}
\begin{tabular}{|l|l|l|l|l|l|}
 \hline
Session id &	Bin 1 &	Bin 2 &	... &	Bin 40 & Label\\ \hline
$$0001$$ &	2 &	0 &	... &	1 & seeker\\
...	& ... & ... & ... & ... & seeker\\
$$0318$$ &	0 &	1 &	... &	0 & avoider \\ \hline
\end{tabular}
\end{table}

\subsubsection{Measures and Analysis}
We used decision tree regressors as well as random forest regressors as models. 
Decision trees are predictive models with a tree-like structure \cite{quinlan1986induction}. The regressors model the relationship of input data within the branches to the class labels in the leaves. During training the data is inserted into the root node. From there, it travels to the leaves though the branches. The branches contain conjunctions of data feature restrictions which redirect the instances into child nodes. The leaves contain the mean values of the training data's class labels. We chose decision trees mainly because they can be visualized and give a good intuition on which features are most relevant. We used regression instead of straight forward classification in order to better incorporate the continuous transition of user types. In addition regressors can be evaluated by receiver operator characteristic (ROC) curves which allow for a more detailed evaluation.
Random forests are ensembles of decision trees \cite{ho1995random}. In order to train a random forest several decision trees are trained on randomized subsets of the data. During training time the data features they are allowed to put restrictions on are also randomly limited. This randomization decorrelates the trees from each other while preserving their ability to generalize across large parts of the data set. During prediction, all trees within the forest vote on the instance\textquotesingle s value. However, regressors model continuous relationships. In order to extract classes from regressors, the continuous space must be divided into sections representing classes via thresholds. We determine these thresholds by evaluating the regressors on a ROC curve. A ROC curve considers different thresholds on which a regressor's predictions could be split into two classes and plots the amount of true positives (correctly labeled seekers) against the amount of false positives (sessions that were predicted to be seekers but are actually avoiders) for each of these thresholds. We then choose the threshold in order to maximize the ratio of true positives to false positives. 


We set the bin size to 15 s for the analysis. We trained the random forest classifier with 200 decision trees. The maximal tree depth was not limited.
When looking at the labels assigned to the participants, we found that we had classified 201 participants as seekers and only 59 participants as avoiders. To adapt the model, we weighed ideator type classes by their frequency in the dataset\footnote{For the code and data, see \url{https://osf.io/7wjya/?view_only=0ba9e138d22e414abd8b868ed594e93e}}.
The data consists of input values of the bins which are the sum of inspiration requests within the bin's time period. For our experiments we created sub-data sets which consist of the first x minutes of the bins but exclude the rest. These sub-data sets are called x-minute data sets.
With these parameters we conducted two experiments: The first experiment trains and predicts scores on session A. We train the random forest on x-minute data sets of on a subset of users, the train data set. Then we score the seekers and avoiders with the regressors on x-minute data sets of the rest of users, the test data sets. We create ROC-curves to determine optimal thresholds with which we classify the test data sets. The metrics we use to evaluate the configurations are defined as follows:
\begin{equation*}
\begin{split}
\text{Accuracy: }&\frac{TP+TN}{TP+FP+FN+TN}\\
\text{Precision: }&\frac{TP}{TP+FP}\\
\text{Recall: }&\frac{TP}{TP+FN}\\
\end{split}
\end{equation*}
Then we visualize one of the trees in order to understand qualitatively how the predictions were made.
In the second experiment the classifiers are trained on session A and predict on session B, thus, generalizing across challenges. The random forest is trained on x-minute data sets of all instances in session A and predicts the classes of all instances on session B. Random forests take the same length inputs that they were trained on, which means that the last 5 min of the second session's bins are not considered as input at any point in the experiment. Otherwise the evaluation is analogous to the first experiment.

\subsubsection{Results}
\begin{figure}[htb]
\centering
\begin{subfigure}[c]{0.52\linewidth}
\includegraphics[width=\linewidth]{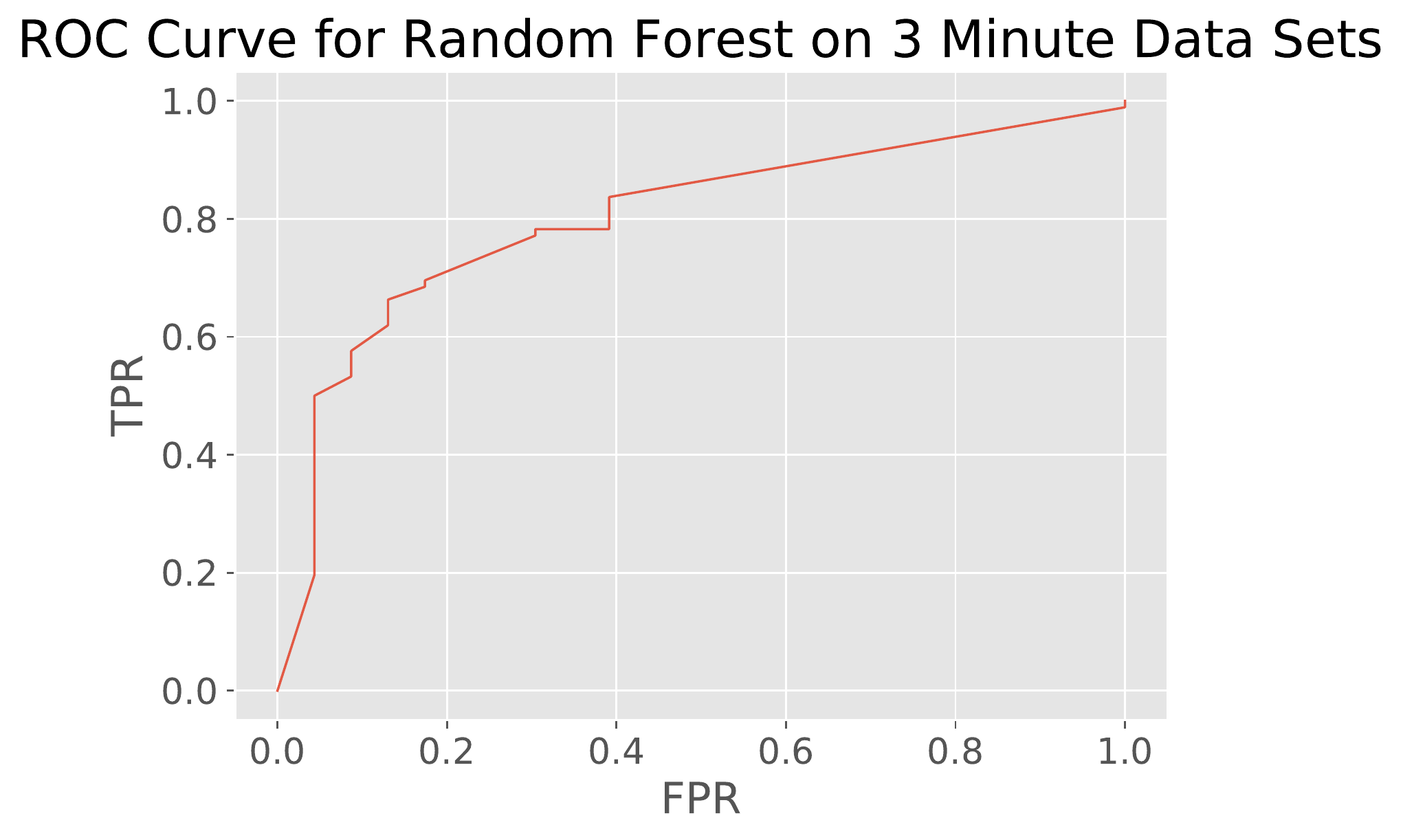}
\subcaption{ROC curve on a 3 minute data set for experiment 1.}
\label{fig:roc-curve-example}
\end{subfigure}
\centering
\begin{subfigure}[c]{0.47\linewidth}
\includegraphics[width=\linewidth]{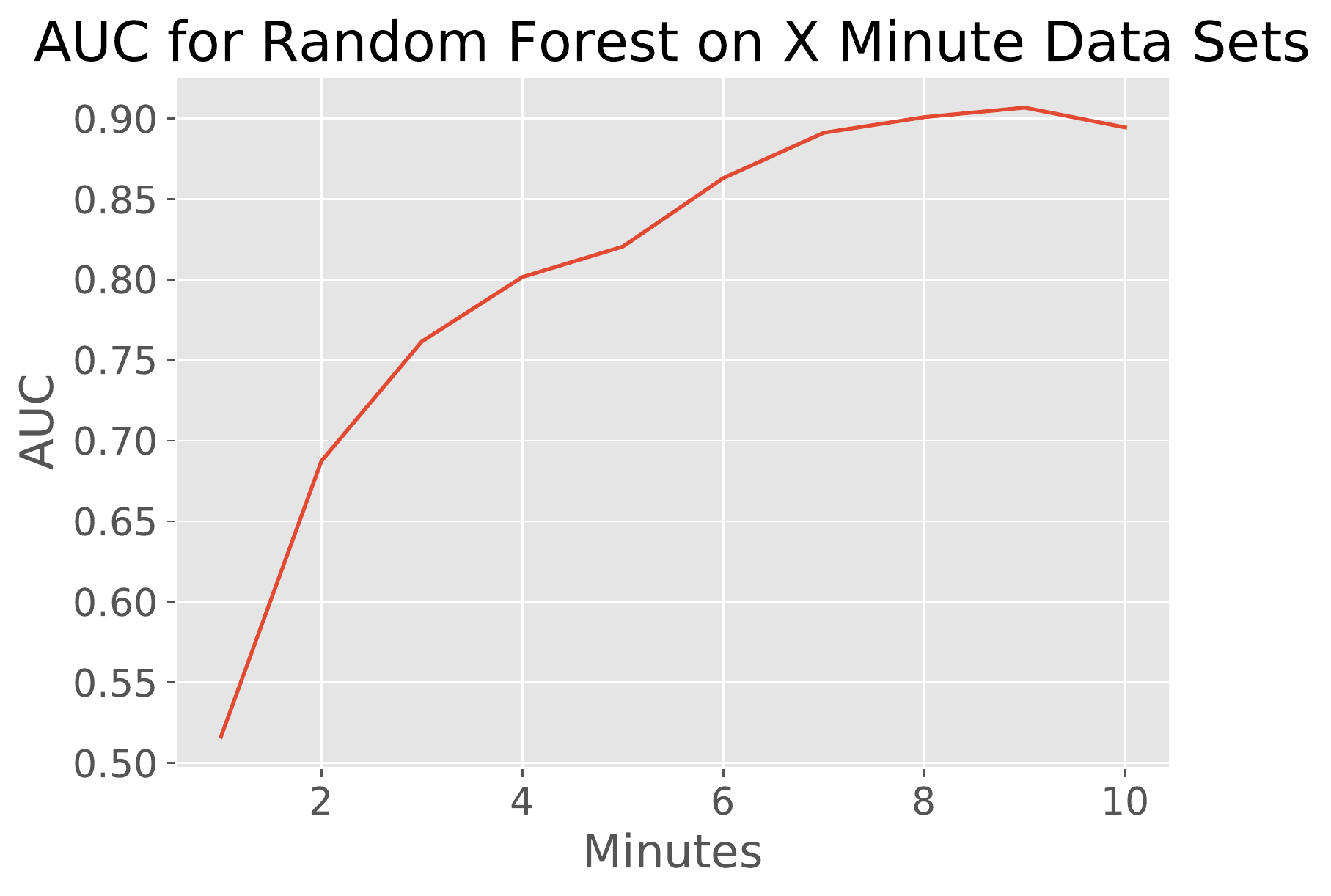}
\subcaption{AUC plot for experiment 1 summarizing the ROC curves.}
\label{fig:auc-exp-1}
\end{subfigure}
\caption{receiver operator characteristic (ROC) and area under the curve (AUC) examples for experiment 1}
\end{figure}

\begin{figure}[htb]
  \centering
  \includegraphics[width=0.7\linewidth]{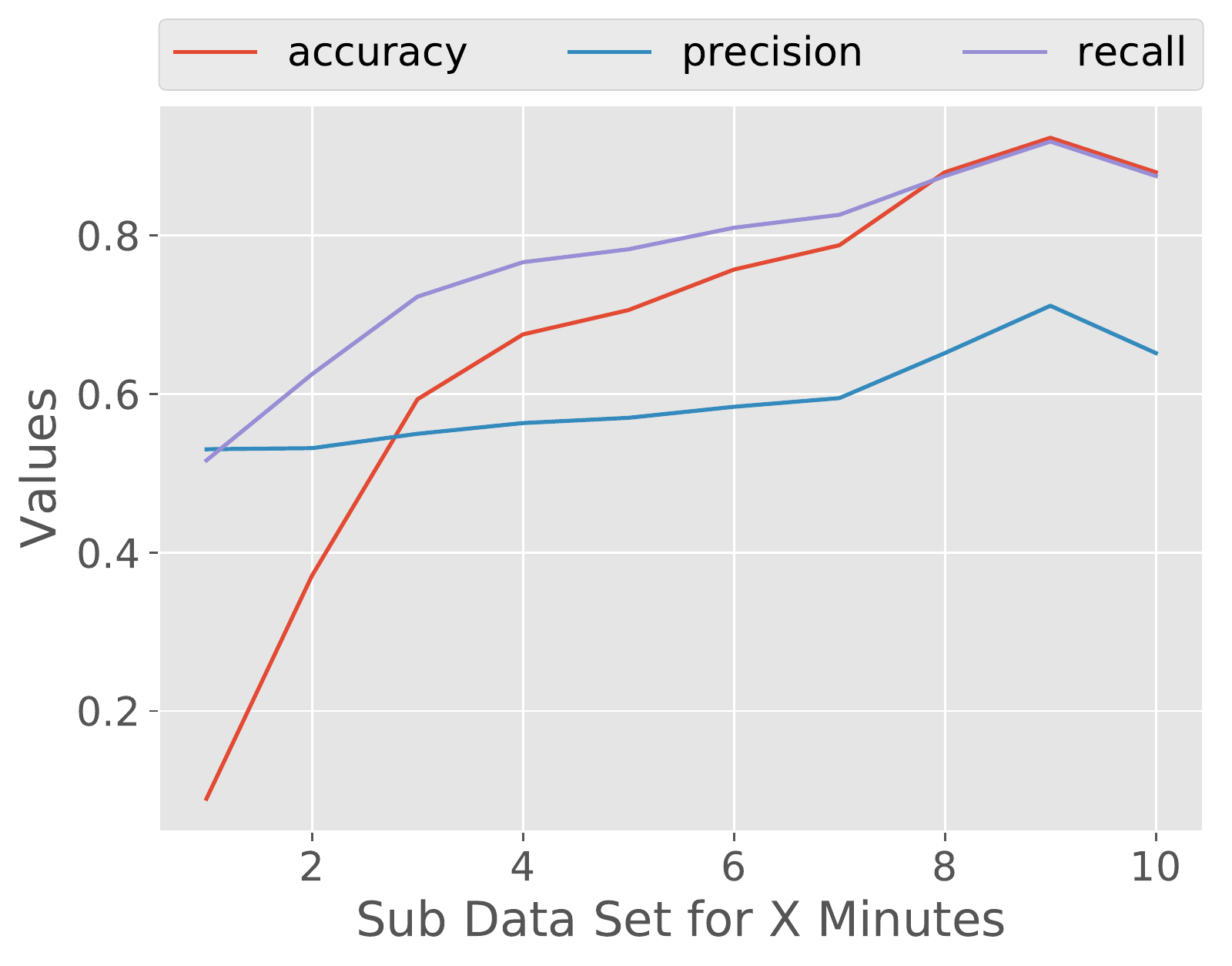}
  \caption{Accuracy, precision and recall for session A ideator type predictions based on a decision tree trained on data sets from session A.}
  \label{fig:ex1-metrics}
\end{figure}

\begin{figure}[htb]
  \centering
  \includegraphics[width=1\linewidth]{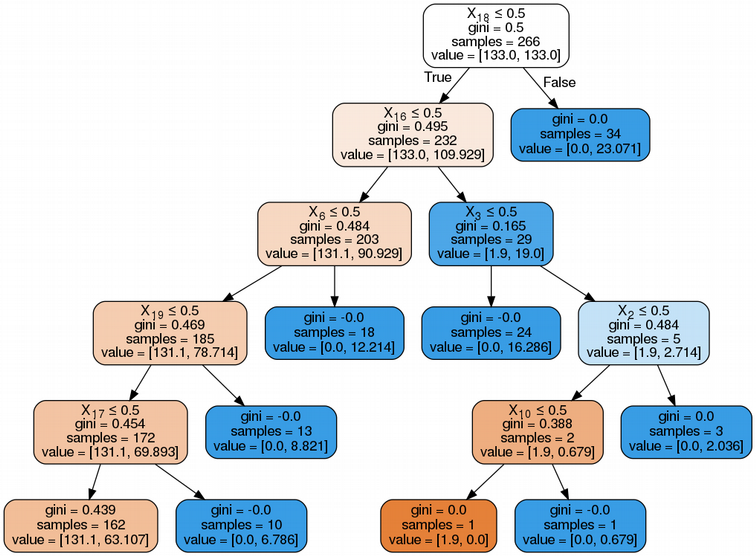}
  \caption{Visualization of a 5-minute decision tree generated by the random forest classifier for experiment 1.}
  \label{fig:tree-ex1-5min}
\end{figure}

\begin{figure}[htb]
  \centering
  \includegraphics[width=0.7\linewidth]{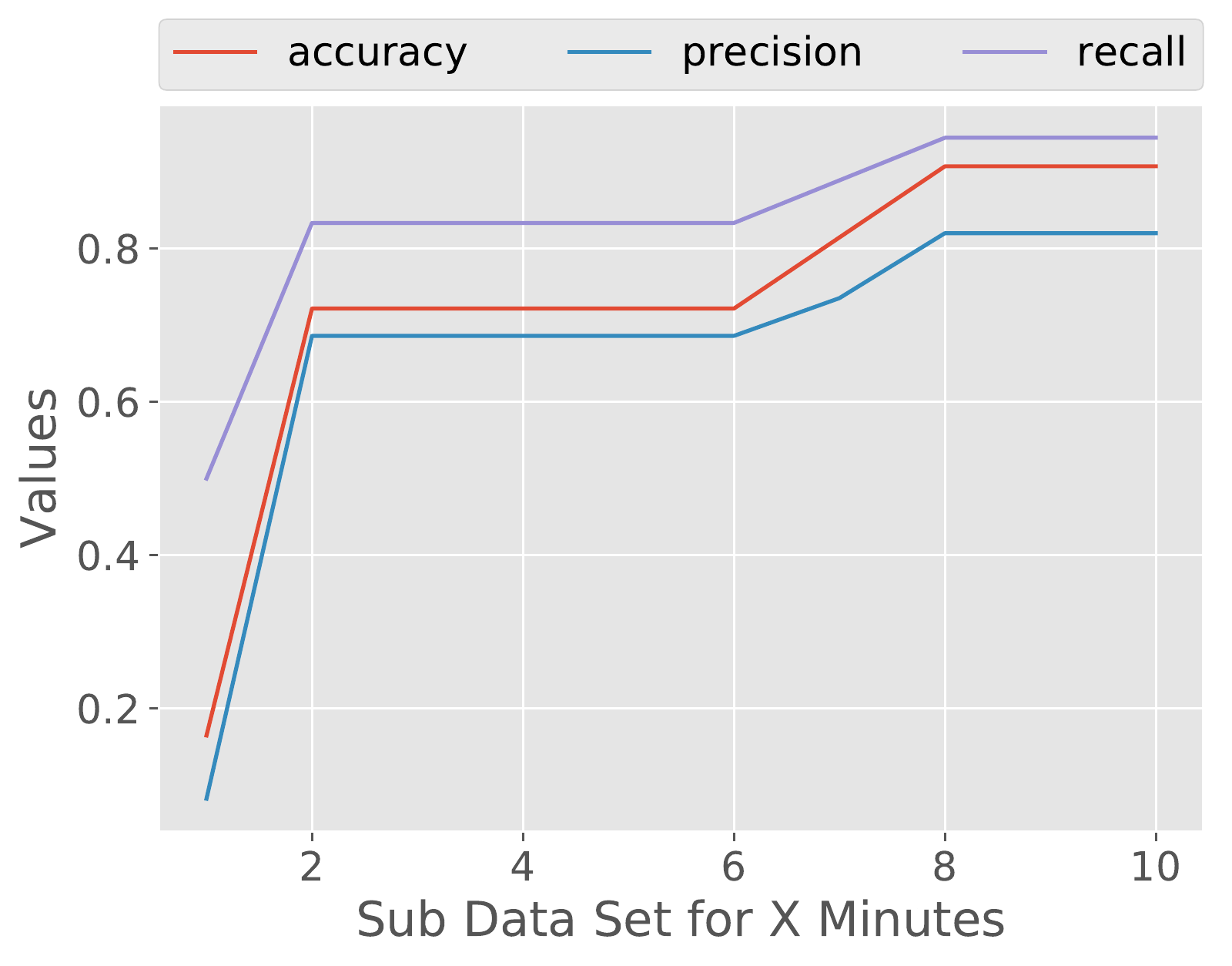}
  \caption{Accuracy, precision and recall for session B ideator type predictions based on a decision tree trained on data sets from session A.}
  \label{fig:ex2-metrics}
\end{figure}

For the first experiment, we created one ROC curve per minute (cf Figure \ref{fig:roc-curve-example} for an example ROC curve at 3 min). We then calculated the area under the curve (AUC) for each ROC curve (Figure \ref{fig:auc-exp-1} shows the result). The AUC increases steadily but most strongly for the first 3-4 minutes then the increase ebbs off quickly. Having determined the thresholds, we calculate accuracy, precision and recall for the prediction with an x-minute input data set. 
Results for accuracy, precision and recall for the prediction results using the session A data are shown in Figure \ref{fig:ex1-metrics}. The increase in predictive power is largest in the first minutes and declines subsequently.
We visualize an individual decision tree for a more qualitative analysis (see Figure \ref{fig:tree-ex1-5min}). Every bin in the data concerns 15 s, for example, X1-X4 contains minute 1. The decisions made close to the tree's root concern more relevant bins than the decisions made closer to the leaves. The computed tree shows that more fundamental decisions focus on the latest available minutes, if the tree uses the first 5 min, then the bins of the fifth minute are most relevant. The same is true for all 10 min.
For the second experiment, we also calculated the ROC curves, AUC, thresholds, and metrics for each one to ten minutes. Figure \ref{fig:ex2-metrics} shows accuracy, precision and recall plotted for x-minute data sets. As the prediction is made on session B, while being trained on session A data, the last 5 minutes of session B are not accounted for by the classifier. Similar to the results of Figure \ref{fig:ex1-metrics}, the predictive power is greatest at the start. Due to the small number of samples in session B, we see no increase in prediction quality for 2 to 6 min.



\section{Discussion}
In the following, we discuss the results of our crowdsourced ideation sessions in the context of our research questions which were stated above (cf Section \ref{sec:research-questions}).

\subsection{Relation between participant and ideator type}
When discussing the insights from the co-located brainstorming sessions, we speculated whether the ideator type either is a character trait or depends on the provided challenge. The latter makes sense insofar that some challenges might be too difficult and others uninteresting for some participants, which might influence their ideation behavior.
We elaborated this question by three analyses. First, we analyzed inspiration requests in the \textit{on-demand} condition. As expected, participants assigned to the group of seekers requested significantly more inspirations. This aligns with our intuition, derived from the pre-study in a co-located setting, that there are individual preferences regarding inspirations. However, the relationship between a person and the provided challenge seems to be more complex than expected: Three participants changed their group membership from avoiders to seekers in the second session. In that study, we presented the participants with two similar, rather technical challenges. Our insights from this condition are inconclusive. In future work, we plan to investigate whether the assigned ideator type remains stable even when two or more fundamentally different challenges (e.g. technical, social, ludic) are compared.
Additionally, we analyzed the accuracy of the between-session predictions of our random forest classifiers and found that, while the accuracy initially increased in the predictive power (similar to the within-challenge accuracy), the cross-domain setting shows no relevant improvement after 3 to 4 min of inspiration requests. In future research, we plan to elaborate on the more complex relationship between ideator type and the challenge provided, for example by explicit elicitation of confidence or challenge comprehension.

\subsection{No significant differences in fluency between seekers and avoiders}
Inspired by the insights from the co-located setting, we supposed that avoiders have a better understanding of their idea generation strategies, thus, might generate more ideas. However, when differentiating study participants based on our heuristic, we were not able to find significant differences in fluency between ideator types. Even though both participant groups use inspirations differently, no differences can be seen by only looking at the fluency. We, therefore, assume that inspirations have a positive effect on ideation outcome but do not have the same impact on all users. More research is needed to better understand the effects of inspirations on different user groups. 
Furthermore, in future research we plan to refine the user model, to include other factors that could potentially outweigh the seeker / avoider classification such as task motivation, task comprehension, confidence and prior experience in brainstorming. We plan to elicit these data based on explicit user preferences provided by the study participants as has already been used in recommendation and information filtering applications (e.g.~\cite{knijnenburg2011each}).

\subsection{Avoiders are negatively impacted by inspiration availability}
Based on the crowdsourced ideation session, we learnt that, when providing avoiders with the ability to request inspirations, it impacted the novelty of their ideas negatively. This effect confirms our initial intuition that avoiders have their own set of strategies when generating ideas. This observation contrasts seekers, who use inspirations as a kind of guideline or scaffold for both understanding the challenge better and becoming inspired by topics, entities or activities in the inspiration provided. This effect, however, shows that there is a distinction between seekers and avoiders that is relevant for providing inspirations. We plan to further investigate this distinction by conducting research in which we use the heuristic approach to adapt the inspirations available in a timely manner (e.g. by disabling or fading out the inspiration button). We assume that this could further help avoiders in leveraging their own idea generation strategies, while at the same time supporting seekers who need inspirations during ideation.

\subsection{Random forest classifiers are promising as a predictor for avoiders / seekers}
We employed random forest classifiers to predict the user's type. The classifier worked especially well in the first few minutes of the ideation session. The accuracy plots show a quick increase in predictive power for both avoiders and seekers and the ROC-curves prove that the increase is due to structures within the user-behavior and not random chance. The increase in predictive power is largest in the first few minutes and declines later, although it never ebbs off entirely. The increase is in contrast to the decision tree visualized, that assigns later minutes a greater predictive power. This effect can be explained because an avoider has a higher probability of trying the inspiration button in the early minutes without using it later whereas it is unlikely in later minutes. 
The random forests were used because they are reliable in dealing with noisy data, achieving comparatively accurate results and finding difficult patterns~\cite{hastie2009random}. In future research, other classifiers more suited to the recursive list-like nature of the data, such as recurrent neural networks or bayesian networks might be more suitable when finding avoiders and seekers on cross-domain settings and different time scales.


\subsection{Limitations}

One potential limitation of the study is the participants' understanding of the procedure and the task of the study. Although the study included an introduction and a short tutorial of the interface, it is not clear that all avoiders deliberately chose not to use inspirations or whether they might not have understood the inspiration mechanism. Future work should include an interactive tutorial to ensure that participants understand the mechanism.
At the current state, the classification into seekers and avoiders is only done by the analysis of inspiration requests. Future work could compare this inspiration request-based model with self-assessment (both prior and after the session) of the participants preferences regarding inspiration.


\section{Conclusion}
Research has shown that the creative outcome of large-scale ideation can be improved by showing inspirations to ideators. However, existing approaches neglect individual preferences.
In the research presented here, we sought to close this gap, and investigated how individual preferences can inform a user model for personalized adaptive inspirations. We conducted our research in two stages. Firstly, we conducted co-located brainstorming sessions which informed an exploratory data analysis. The insights collected in these studies finally informed a user model. Using this model, we differentiate ideation participants into seekers, i.e. users who appreciate inspirations, and avoiders, i.e. users which are distracted by inspirations and dislike them. We validated this user model in online experiments.
In the latter, we investigated how stable the assignment of people to one ideator type is. We, furthermore, evaluated the impact of the type and the availability of inspiration on ideation metrics (e.g. fluency). Lastly, we trained and deployed a random forest approach to test how accurately we can predict ideator types with incomplete session information.
Our results show that, in terms of numbers of ideas generated, there is no significant differences between ideator types. However, when looking at the ideas rated most novel, we found that while seekers were influenced positively by the availability of inspiration, avoiders' highest novelty decreased with inspiration availability. When using the random forest-based approach for classifying participants, we found that the first three minutes of a session provided non-linear information gain for the predictor. After a time of three minutes, we were able to classify 73\% of the ideator types correctly.
These findings show that ideator types provide relevant information when designing adaptive inspirations within large-scale ideation systems. Furthermore, random forests provide a promising heuristic to determine ideator types based on incomplete session information.
However, our research has a number of limitations. The relationship between ideator type and challenge proved more complex than anticipated. In future work, we will approach this by evaluating the impact of thematically diverse challenges. Moreover, we focused on the principal differences between ideator types by using static conditions. In future work we will therefore evaluate the impact of dynamically adding or removing possibilities to request inspirations.

\begin{acks}
This work is supported by the German Federal Ministry of Education and Research, grant 01IO1617 (“Ideas to Market”).
\end{acks}

\bibliographystyle{ACM-Reference-Format}
\bibliography{references}


\end{document}